---------------------------------------------------------------------------
############################## LaTex Copy #################################
---------------------------------------------------------------------------

\documentstyle[12pt]{article}

\newcommand{\ap}[4]{{\em #1} - Ann.~Phys.               {\bf #2},  #3  (19#4)}

\newcommand{\jp}[4]{{\em #1} - J.~Phys.                 {\bf #2},  #3  (19#4)}

\newcommand{\mpl}[4]{{\em #1} - Mod.~Phys.~Lett.        {\bf #2},  #3  (19#4)}

\newcommand{\np}[4]{{\em #1} - Nucl.~Phys.              {\bf #2},  #3  (19#4)}

\newcommand{\nc}[4]{{\em #1} - Nuovo Cim.               {\bf #2},  #3  (19#4)}

\newcommand{\pr}[4]{{\em #1} - Phys.~Rev.               {\bf #2},  #3  (19#4)}
\newcommand{\prl}[4]{{\em #1} - Phys.~Rev.~Lett.        {\bf #2},  #3  (19#4)}

\newcommand{\ptp}[4]{{\em #1} - Progr.~Theor.~Phys.     {\bf #2},  #3  (19#4)}

\newcommand{\bel}[1]{\begin{equation}\label{#1}}
\newcommand{\be}{\begin{equation}}
\newcommand{\ee}{\end{equation}}

\newcommand{\beal}[1]{\begin{eqnarray}\label{#1}}
\newcommand{\bea}{\begin{eqnarray}}
\newcommand{\eea}{\end{eqnarray}}

\newcommand{\bean}{\begin{eqnarray*}}
\newcommand{\eean}{\end{eqnarray*}}

\newcommand{\ba}{\begin{array}}
\newcommand{\ea}{\end{array}}

\newcommand{\dv}{\partial}

\newcommand{\bamd}[4]{\left( \begin{array}{cc}{#1}&{#2}\\
{#3}&{#4}\end{array} \right)}

\newcommand{\bamq}[4]{\left( \begin{array}{cccc}{#1}&{#2}&{#3}&{#4}\\}
\newcommand{\bamc}[5]{\left( \begin{array}{ccccc}{#1}&{#2}&{#3}&{#4}&{#5}\\}
\newcommand{\eam}{\end{array} \right)}

\newcommand{\raw}{\rightarrow}

\newcommand{\cg}{\gamma}
\newcommand{\dg}{\delta}

\newcommand{\pg}{\phi}

\newcommand{\Lb}{\Large \bf}

\newcommand{\fs}{\footnotesize}

\newcommand{\bii}{\begin{itemize}}
\newcommand{\eii}{\end{itemize}}

\newcommand{\ben}{\begin{enumerate}}
\newcommand{\een}{\end{enumerate}}

\newcommand{\bq}{\begin{quote}}
\newcommand{\eq}{\end{quote}}

\newcommand{\bc}{\begin{center}}
\newcommand{\ec}{\end{center}}

\newcommand{\bt}{\begin{tabular}}
\newcommand{\et}{\end{tabular}}

\newcommand{\br}{\begin{flushright}}
\newcommand{\er}{\end{flushright}}

\newcommand{\bl}{\begin{flushleft}}
\newcommand{\el}{\end{flushleft}}

\newcommand{\f}[1]{\footnote{#1}}

\newcommand{\vs}[1]{\vspace*{#1}}

\newcommand{\new}{\pagebreak}

\newcommand{\bb}{\begin{thebibliography}{99}}
\newcommand{\eb}{\end{thebibliography}}
\newcommand{\bi}{\bibitem}

\newcommand{\btp}{\begin{titlepage}}
\newcommand{\etp}{\end{titlepage}}

\newcommand{\con}{\section{Conclusions}}

\newcommand{\ddg}{\dg \pg_{i}}
\newcommand{\bdg}{\dg \pg_{i}^{+}}

\newcommand{\ddl}{\dg {\cal L}}

\begin{document}

\hyphenation{mech-ani-cs commu-tes dire-cti-on negati-ve-ly further-mo-re}

\btp

\bc
September, 1995

\vs{3cm}

{\Lb The Quaternionic Dirac Lagrangian}

\vs{2cm}

{\sc Stefano De Leo}$^{ \; a)}$ and {\sc Pietro Rotelli}$^{ \; b)}$ \\

\vs{1cm}

{\it Universit\`a  di Lecce, Dipartimento di Fisica\\
Instituto di Fisica Nucleare, Sezione di Lecce\\
Lecce, 73100, ITALY}

\vs{2cm}

{\bf Abstract}

\ec
We discuss the use of the variational principle within quaternionic quantum
mechanics. This is non-trivial because of the non commutative nature of
quaternions. We derive the Dirac Lagrangian density corresponding to the
two-component Dirac equation. This Lagrangian is complex projected as
anticipated in previous articles and this feature is necessary even for a
classical real Lagrangian.

\vs{2cm}

\noindent{\fs a) e-mail:} {\fs \sl deleos@le.infn.it}\\
{\fs b) e-mail:} {\fs \sl rotelli@le.infn.it}

\etp

\new

\section{The variational principle}
The standard derivation of the Euler-Lagrange field equations implicitly
assumes that the variation in the action $I$ of a given field $\ddg$
may be commuted to, say, the extreme right. Thus after a functional
integration by parts, and neglecting the surface terms one obtains the
field equations,
\bel{one}
\frac{\dv {\cal L}(x)}{\dv \pg_{i}(x)}=
\dv_{\mu}\frac{\dv {\cal L}(x)}{\dv_{\mu} \pg_{i}(x)} \; \; ,
\ee
where $\cal L$ is the Lagrangian density and $x$ represents space-time.

Somewhat surprisingly this form of the field equations survives in
field theory, that is after second quantization. The reason that this is
not a priori obvious is because when the fields become operators and
non-commutating, the translation of $\ddg$ is not necessarily without
consequences. Indeed, we should be obliged to define with attention the
significance of the functional derivatives in eq.~(\ref{one}). This
fact does not in practice produce difficulties as a consequence of:
\ben
\item For bilinear terms (e.g. mass or kinetic energy) the variations naturally
lie on the right (for $\ddg$) or the left (for $\bdg$);
\item In any case some authors implicitly assume that, when $\ddg$ is bosonic
it commutes with everything;
\item When $\pg_{i}$ is a fermion field $\ddg$ is assumed at least to
commute with any bosonic fields and with $\pg_{i}$ itself. This last fact
(by no means obvious) leads to a {\em natural} extraction of $\ddg$ to the
right or the left.
\een
More specifically, within normal ordered products there always exists a
natural direction of extract given the fact that all creation operators (
bosonic or fermionic) commute amongst themselves.

It is however an assumption that the variation in $\pg_{i}$, i.e. $\ddg$
has the same characteristic commutation relations of $\pg_{i}$. These
implicit assumptions are brought to the fore when one considers
quaternionic fields both in classical and quantum field theory. For it is
unthinkable that a quaternionic $\ddg$ created within a Lagrangian density
can, in general, be commuted without consequence to the left or the right.
The example of the free Dirac Lagrangian treated in this paper will demonstrate
some of the difficulties.

Before entering explicitly into the world of quaternionic fields we wish to
discuss the limitations, if any, applied to $\ddg$ in standard (complex)
quantum mechanics. Our objective, which will have an application in this
paper, is to demonstrate that the
properties of $\ddg$ may be substantially divers from
those of $\pg_{i}$. For example, we shall observe first that even if
the classical Lagrangian density is necessarily real, $\ddg$ or more correctly
$\ddl$ may be formally complex (or even quaternionic) since at the
end $\ddl\equiv 0$ to obtain the Euler-Lagrange equations. It is true
that in general we may limit our attention to a subset of $\ddg$
($\ddl$) without losing the Euler-Lagrange equations, so that our
observation seems accademic, but it is not without consequences with
quaternions as we shall see.

Consider one of the simplest of all particle Lagrangian densities, that for
two classical scalar fields ($\pg_{i}$-real {\fs $i=1, \; 2$}) without
interactions:
\bea
{\cal L} &    =   & \frac{1}{2} \dv_{\mu} \pg_{1} \dv^{\mu} \pg_{1} -
\frac{m^{2}}{2} \pg_{1}^{\; 2} +
\frac{1}{2} \dv_{\mu} \pg_{2} \dv^{\mu} \pg_{2} -
\frac{m^{2}}{2} \pg_{2}^{\; 2} \nonumber \\
         & \equiv & \dv_{\mu} \pg^{+} \dv^{\mu} \pg -
m^{2} \pg^{+} \pg
\eea
where
\[ \pg \equiv \frac{1}{\sqrt{2}} ( \pg_{1} + i \pg_{2} ) \; \; ,\]
and
\[ \pg^{+} \equiv \frac{1}{\sqrt{2}} ( \pg_{1} - i \pg_{2} ) \; \; .\]
The well known corresponding Euler-Lagrange equations are:
\be
(\dv_{\mu} \dv^{\mu} + m^{2}) \pg_{i} = 0 \; \; ,
\ee
\bc
{\fs ($i = 1, \; 2$) ,}
\ec
or, equivalently,
\be
(\dv_{\mu} \dv^{\mu} + m^{2}) \pg = 0 \; \; .
\ee
Now to obtain ``directly'' the last equation one performs the very particular
variation of $\pg$ ($\pg^{+}$)
\beal{five}
\pg       & \raw & \pg \nonumber \\
\pg^{+}   & \raw & \pg^{+} + \dg \pg^{+}
\eea
i.e. in order to obtain the corresponding Euler-Lagrangian equation one
treats $\pg$ and $\pg^{+}$ as {\em independent} fields. In second
quantization these fields indeed contain independent creation and
annihilation operators corresponding to positive and negatively charged
particles\f{Actually the ``electric'' charge $e$ of the fields depends upon
whether the global gauge group is to be made local or not. This fact is a
choice {\em not} determined by the free Lagrangian.}. To satisfy
eq.~(\ref{five}) we must necessarily have,
\be
\dg \pg_{1} + i \dg \pg_{2} = 0
\ee
and this means, that the {\em variations} in the originally real $\pg_{i}$
fields are complex (if $\dg \pg_{1}$ is real then $\dg \pg_{2}$ is pure
imaginary etc.).

Of course we could obtain the equivalent result from variations separately of
$\pg_{1}$ and $\pg_{2}$, e.g. varying $\pg_{1}$ (with $\pg_{2}$ constant)
\bea
\pg       & \raw & \pg + \dg \pg_{1} \nonumber \\
\pg^{+} & \raw & \pg^{+} + \dg \pg_{1}
\eea
yielding, after a double integration by parts
\be
(\dv_{\mu} \dv^{\mu} + m^{2}) (\pg + \pg^{+}) = 0
\ee
and the corresponding result for $\pg_{2}$. Thus while not
obligatory we can in this latter approach consider only
{\em real} $\ddg$. However, there is a subtle difference in the two
approaches, which readily passes unobserved. In the case when only
$\pg_{1}$ or $\pg_{2}$ is varied the $\ddg$ appears both to the left and to
the right. Only if $\ddg$ commutes with $\pg_{i}$ are the two methods
equivalent.

The point of these observations is that it is possible to impose diverse
conditions on the Lagrangian density, on the fields, and on their variations.
This is ignored in standard quantum mechanics but is important for
what follows.

\section{The two component Dirac Equation}
The adoption of quaternions as the base number system in quantum mechanics
allows one to define a quaternionic Dirac equation~\cite{rot} in which
the wave function is characterised by having only two components
(consequently the Schr\"odinger-Pauli equation applies to a one-component wave
function). This follows from the fact that the Dirac algebra upon the reals
(but not upon the complex) has a two dimensional irreducible representation
with quaternions. The standard $4\times 4$ complex gamma matrices are in
fact reducible. This structure is consistent if the momentum operator
$p^{\mu}$ is defined by
\be
p^{\mu} = \dv^{\mu} \mid i
\ee
where a {\em barred} operator $A\mid b$ acts upon a general wave function
$\psi$ by
\be
(A\mid b)\psi \equiv A\psi b \; \; .
\ee
In general $A$ will be a matrix and $b$ a complex ${\cal C}(1, \; i)$
number, where $i$ is one of the imaginary $(i, \; j, \; k)$ units of a
quaternion:
\be
q=a_{0}+a_{1}i+a_{2}j+a_{3}k \; \; ,
\ee
with
\[i^{2}=j^{2}=k^{2}=-1 \; \; ,\]
\[ij=-ji=k \; \; \; \; ( \; cyclic \; ) \; \; ,\]
\[ a_{0, \; 1, \; 2, \; 3} \in {\cal R} \; \; .\]
The quaternionic conjugation is defined by
\be
q^{+}=a_{0}-a_{1}i-a_{2}j-a_{3}k \; \; .
\ee

The Dirac equation then reads in covariant form
\bel{twelve}
\cg_{\mu}\dv^{\mu} \psi i - m\psi = 0
\ee
with as a possible choice of the $\cg_{\mu}$
\bea
\cg_{0}=\bamd{1}{0}{0}{-1} & , & \cg_{1}=i\bamd{0}{1}{1}{0}
\; \; , \nonumber \\
\cg_{2}=j\bamd{0}{1}{1}{0} & , & \cg_{3}=k\bamd{0}{1}{1}{0} \; \; .
\eea
This equation has four plane wave ($\propto e^{-ipx}$) solutions which
correspond to those of the standard Dirac equation if one adopts the
complex scalar product (complex geometry~\cite{rem})
\be
<\psi \mid \pg >_{\cal C} \; = \frac{1}{2} <\psi \mid \pg > -
\; \frac{i}{2} <\psi \mid \pg > i
\ee
where $<\psi \mid \pg >$ is the quaternionic scalar product,
\be
<\psi \mid \pg > = \int \psi^{+} \pg \; d \tau
\ee
so that the four plane wave solutions are (complex) orthogonal.
Actually the need of
this scalar product is anticipated by our choice of $p^{\mu}$ which is not
hermitian on a quaternionic geometry. The complex scalar product
was first introduced in the definition of tensor products in quaternionic
quantum mechanics~\cite{hor}, and appears essential for the ``translation''
of complex quantum mechanics to a quaternionic version~\cite{del1}.

Our main objective in this work is to derive the Dirac Lagrangian (density)
which yields eq.~(\ref{twelve}), and this will be done in the next section.
We conclude this section with some notes and comments upon this use of
quaternions in quantum mechanics.
\ben
\item The objective of reproducing almost all standard results in quantum
mechanics is achieved;
\item There are differences within the bosonic sector. For example there is
a doubling of solutions in the Klein-Gordon equation~\cite{del2};
\item There exist however projected equations with only the standard number
of solutions~\cite{del3a,del3b};
\item A particular example of point 3 is the {\em non} equivalence (with
quaternions) of the Maxwell equation with the Duffin-Kemmer-Petiau
equation. The latter has no doubling of solutions, and may be identified
with a projected Maxwell equation;
\item It is possible to invent new quaternionic equations equivalent to
pairs of complex equations in the same way as the Schr\"odinger equation
can be rewritten as a pair of real equations;
\item There exists a not yet completely investigated generalization of group
theory or more precisely of representation theory.
\een

Finally, we are always intrigued by the fact that had Schr\"odinger
considered quaternionic solutions to his equation, he would have found two
(within a complex geometry) and probably discovered the existence of spin.

\section{The Dirac Lagrangian}
Let us consider, as a first hypothesis, the traditional form for the Dirac
Lagrangian density:
\bel{lag}
{\cal L}=\frac{i}{2}[\bar{\psi}\cg^{\mu}(\dv_{\mu}\psi)-
(\dv_{\mu}\bar{\psi})\cg^{\mu}\psi]-m\bar{\psi} \psi
\ee
as given for example by Itzykson and Zuber~\cite{itz}. The
position of the imaginary unit is purely
conventional in~(\ref{lag}) but with a quaternionic
number field we must recognize that the $\dv_{\mu}$ operator is more
precisely part of the first quantized momentum
operator $\dv_{\mu}\mid i$ and that hence
only the hermitian conjugate part of the kinetic term (second half of the
bracket expression in eq.~(\ref{lag})) appears with the imaginary unit in
the ``correct'' position. Thus the correct form of the kinetic term reads:
\bel{lagk}
{\cal L}_{K}=\frac{1}{2}[\bar{\psi}\cg^{\mu}\dv_{\mu}\psi i-i
(\dv_{\mu}\bar{\psi})\cg^{\mu}\psi] \; \; .
\ee
We observe that
this modification of eq.~(\ref{lag}) is also justified by the simple
requirement that $\cal L$ be hermitian. We could
of course multiply eq.~(\ref{lagk}) by the ``hermitian''
$-i\vert i$ which inverts the order of the imaginary unit, but by
integrating by parts this may be reformulated as in eq.~(\ref{lagk}).
Notice that we cannot change the sign of $\cal L$ without changing the
sign of the Hamiltonian $\cal H$.

The requirement of hermiticity however says nothing about the Dirac mass
term in eq.~(\ref{lag}). It is here that appeal to the variational
principle must be made. A variation $\dg \psi$ in $\psi$ cannot in
eq.~(\ref{lagk}) be brought to the extreme right because of the imaginary unit
in the first half of the expression. The only consistent procedure is to
generalize the variational rule that says that $\psi$ and $\bar{\psi}$
must be varied
{\em independently}. We thus apply independent variations to $\psi$
($\dg \psi$) and $\psi i$ ($\dg (\psi i)$). Similarly for $\dg \bar{\psi}$ and
$\dg (i \bar{\psi})$. Now to obtain the desired Dirac equation for $\psi$
and its adjoint equation for $\bar{\psi}$ we are obliged to modify the mass
term into
\be
{\cal L}_{m}=-\frac{m}{2}[\bar{\psi} \psi-i\bar{\psi} \psi i] \; \; .
\ee
The final result for $\cal L$ is
\bel{lagd}
{\cal L}_{D}=\frac{1}{2}[\bar{\psi}\cg^{\mu}\dv_{\mu}\psi i - i
(\dv_{\mu}\bar{\psi})\cg^{\mu}\psi]-
\frac{m}{2}[\bar{\psi} \psi-i\bar{\psi} \psi i] \; \; .
\ee
Considering this last equation we observe that it is nothing other than the
complex projection of equation~(\ref{lag})
\bel{com}
{\cal L}_{D}=\frac{1-i\mid i}{2} {\cal L} \; \; .
\ee
Indeed, while $\cal L$ in eq.~(\ref{lag}) is quaternionic and
with the modification of ${\cal L}_{K}$ in eq.~(\ref{lagk}) hermitian, the
form given in eq.~(\ref{lagd}) is purely complex and hermitian.

This observation is even more subtle with classical fields for now $\cal L$
defined by
\bel{lagr}
{\cal L}={\cal L}_{K}-m\bar{\psi}\psi
\ee
is both hermitian and {\em real}. Thus it may be objected that the complex
projection in eq.~(\ref{com}) is superfluous. For $\cal L$
itself this is true but for {\em quaternionic} variations in
the fields $\dg \psi$ etc. a difference exists. The variation $\dg {\cal L}$
of eq.~(\ref{lagr})
is in general quaternionic while, because of eq.~(\ref{com}), the variation
$\dg {\cal L}_{D}$ is necessarily always {\em complex}. Furthermore
eq.~(\ref{lagr}) would not yield the correct field
equation through the variational principle unless we limit $\dg \psi$ to
complex variations notwithstanding the quaternionic nature of the fields.
We consider this latter option unjustified and thus select for the formal
structure of ${\cal L}$ that of eq.~(\ref{lagd}).

We note that as described in the first section, one should as a rule define
the numerical basis of the fields and separately of the variations. For
classical fields we have the case of real fields with complex variations
(charged scalar fields) and in our studies the {\em possibility} of
quaternionic fields but with only complex variations (the second option
described above).

Furthermore one should specify the numerical nature of both $\cal L$ and
$\dg {\cal L}$. For example with our choice given by eq.~(\ref{lagd}) we
have:
\bc
{\fs \bt{ll}
$\psi \; , \; \bar{\psi} \; ,$ & quaternionic fields ; \\
$\dg \psi \; , \; \dg (\psi i) \; , \; \dg \bar{\psi} \; , \;
\dg (i \bar{\psi}) \; ,$ &
quaternionc variations ;\\
${\cal L}_{D} \; ,$ & hermitian, complex projected Lagrangian ;\\
$\dg {\cal L}_{D} \; ,$ & complex but generally non hermitian variation .
\et}
\ec

\section{The invariance groups of ${\cal L}_{D}$}
Having obtained ${\cal L}_{D}$ in the previous section we may ask which if
any group (global) leaves this Lagrangian invariant. Remembering that the
single particle fields are one-component quaternionic functions, the most
natural transformation is,
\bea
\psi       & \raw & f \psi g \nonumber \\
\bar{\psi} & \raw & g^{+} \psi^{+} f^{+} \cg^{0} \; \; .
\eea
Now the quaternionic nature of the $\cg^{\mu}$ matrices limits $f$, and for
arbitrary $\psi$ (since we assume no knowledge at this stage of the field
equations\f{We recall the well known fact that ${\cal L}_{D}$ becomes
identically null, if the field equations are applied to $\psi$ and
$\bar{\psi}$.}) leads to the conclusion that $f$ must be real and commute
with each $\cg^{\mu}$. On the other hand the complex
projection of ${\cal L}_{D}$ permits $g$ to be complex (it
can then be ``extracted'' to the right and then commuted through
${\cal L}_{D}$). Furthermore for invariance,
\be
gg^{+} ff^{+}=\vert g \vert^{2} f^{2} = 1 \; \; .
\ee
We find with our representation for $\cg^{\mu}$ that $f$ is proportional to
the unit matrix, and thus being real its magnitude may be completely
absorbed within $g$, or equivalently $f$ may set by definition to $1$.
Whence, the only invariance group is defined by
\beal{ge}
\vert g \vert^{2} &   =  & 1 \; \; ,\nonumber \\
g  \; \;  \;      &  \in & {\cal C}(1, \; i) \; \; .
\eea
For Lie groups this implies that $g$ is a member of $U(1, \; c)$ the
{\em complex} unitary group.

Remembering that the Glashow group~\cite{gla} for the Weinberg-Salam
theory~\cite{wei} is
$SU(2, \; c)\times U(1, \; c)$ we observe that this $U(1, \; c)$ group may
be identified with ours and that our field must necessarily
be a singlet (scalar)
under $SU(2, \; c)$. Since $\psi$ represents a {\em single} fermion spin
$\frac{1}{2}$ field (the $\cg^{\mu}$ are $2\times 2$ quaternionic matrices)
it is not surprising, and indeed necessary that it represents a singlet
under $SU(2, \; c)$.

The interesting feature is what happens if we select a reducible $4\times 4$
complex set of
$\cg^{\mu}$ matrices. Now the number of the fermion particles is two.
Nevertheless, the existence of a complex $SU(2, \; c)$ group acting
on $\psi$ from the left seems excluded. On the other hand, as we have
described in detail
elsewhere~\cite{del4} the
group $SU(2, \; c)$ is isomorphic (at the generator algebra level) with
$U(1, \; q)$, the unitary quaternionic group with elements
\[ g \sim e^{ia+jb+kc} \]
with
\[ a, \; b , \; c \in {\cal R} \; \; , \]
which is the simplest of all unitary quaternionic groups.

Now it is still not obvious that this group is an invariance group of
${\cal L}_{D}^{(2)}$, where the superscript $^{(2)}$ indicates the presence
of two fermionic fields (with the same mass) since the $\cg^{\mu}$ matrices
are complex, not real. However, it is readily verified
that an equivalent set of $\cg^{\mu}$
matrices (with the same commutation relations) are given if the $i$-factors
in $\cg^{\mu}$ are substituted by $1\mid i$. Whence the imaginary units
appear to the {\em right} of $\psi$ and the elements
given by eq.~(\ref{ge}) commute with $\cg^{\mu}$ and hence cancel within
${\cal L}_{D}$. One may also derive the similarity transformation which
performs the above change of representation and thus derive the
corresponding representation\f{We observe that while our $U(1, \; q)$ group
acts from the {\em left} on $\psi$
and $U(1, \; c)$ acts on the {\em right}, this has nothing to do
with the helicity indices $L$ and $R$ of the Weinberg-Salam theory.} of
$U(1, \; q)$ for our original (conventional) choice of complex $\cg^{\mu}$.

We note that our $U(1, \; q)$ group consists of
quaternionic ``phase'' factors multiplied by the unit $4\times 4$
matrix\f{An additional right acting complex phase $U(1, \; c)$ is also
allowed, but this is
equivalent to that already described in the previous section.}. We may ask
if there are other $4\times 4$ (essentially real) matrices which commute
with the four dimensional Dirac gamma matrices. However there is no other
matrix that commutes with all four gamma matrices so the maximum global
invariance group is indeed $U(1, \; q) \mid U(1, \; c)$ to be identified with
the (complex) Glashow group $SU(2)\times U(1)$.

\con
We begin our conclusions from the end results of the last section. We wish
to recall that not every global invariance group is automatically
gauged into the
corresponding local group in gauge theory. For example, the so-called
complex bosonic Lagrangian may a priori (before eventual gauging) represent
two {\em neutral} equal mass particles or a complex pair of
oppositely charged particle-antiparticle. Thus even in the case of a Dirac
Lagrangian with complex ($4\times 4$) Dirac matrices we do not have
necessarily a doublet under $U(1, \; q)$. So that we have not {\em derived}
the gauge group of the Salam-Weinberg model. The pair
may consist of two singlets. However it was not even obvious a priori that a
global invariance group $U(1, \; q)$ isomorphic at the generator level with
$SU(2, \; c)$ existed. {\em We have thus shown that even with quaternionic
fields, it is possible to impose a Glashow group invariance and that this
occurs by merely adopting reducible gamma matrices}.

Our viewpoint is that $SU(2, \; c)$ invariance in particle physics is really
indication of invariance under the simplest of all unitary quaternionic
groups. We also have a complex invariance group $U(1, \; c)$ but this is
justified by
the complex projection of the Lagrangian density. This complex projection
was {\em not} imposed to obtain the $U(1, \; c)$ factor group but in order
to obtain the desired quaternionic Dirac equation. The fact that this group
exists in nature as the weak-hypercharge group is an undoubted success of
this model.
Further, the automatic appearance of this complex unitary group is expected
whatever the left acting (quaternionic) unitary group is. This strongly
suggests that in the search for grand unified theories one should consider
preferentially a product group of the type $G\mid U(1, \; c)$ with $G$ a
{\em quaternionic} unitary group, conditioned to commute with the Dirac
gamma matrices in the chosen representation.

Let us recall the other results of this paper. We have discussed
the application of the variational principle to Lagrangians with possibly
quaternionic fields. We noted en passant that even within standard field
theory the limitation of variations in the fields to complex variations is
an implicit assumption, since the variations themselves have no physical
content. Thus quaternionic variations have always {\em existed} even
if not traditionally
applied. Of course this observation can be generalized to even more exotic
variations, e.g. supersymmetric variations containing Grassman terms.

Finally, whereas in this work the need of a complex projection of
the Diarc Lagrangian density is demonstrated, for the scalar field Lagrangian
it is an
assumption with interesting consequences\cite{del3a}. We suggest that, for
reasons which still elude us, all terms in the quaternionic Lagrangians
including interaction terms must be complex projected.

\bb

\bi{rot}
\mpl{Rotelli P}{A4}{933}{89}.
\bi{rem}
\jp{Rembieli\`nski J}{A11}{2323}{78}.
\bi{hor}
\ap{Horwitz LP, Biedenharn LC}{157}{432}{84}.
\bi{del1}
\ptp{De Leo S, Rotelli P}{92}{917}{94}.
\bi{del2}
\pr{De Leo S, Rotelli P}{D45}{575}{92}.
\bi{del3a}
{\em De Leo S, Rotelli P} - hepth/9506179 {\fs to appear in
Int.~J.~of Mod.~Phys.~A.}
\bi{del3b}
{\em De Leo S} - hepth/9509??? {\fs to appear in
Prog.~Theor.~Phys.}
\bi{itz}
{\em Itzykson C, Zuber JB} - {\sl Quantum Field Theory} {\fs McGraw-Hill
Book Co, Singapore 1985.}
\bi{gla}
\np{Glashow SL}{22}{579}{61}.
\bi{wei}
\prl{Weinberg S}{19}{1264}{67}.\\
{\em Salam A} - {\sl Proc.~8$^{th}$ Nobel Symposium, Weak and
Electromagnetic Interaction} {\fs ed.~Svartholm, 367 (1968).}
\bi{del4}
\nc{De Leo S, Rotelli P}{B110}{33}{95}.

\eb

\end{document}